\documentclass[preprint]{ptephy_v1}%%%%%% to generate preprint number with ptep logo

\usepackage{hyperref}
\usepackage{amsmath} %for dealing with mathematics,
\usepackage{amsthm} %for dealing with theorem environments,
\usepackage{hyperref} %for linking the cross references
\usepackage{graphics} %for dealing with figures.
\usepackage{algorithmic} %for describing algorithms
\usepackage{url} %It provides better support for handling and breaking URLs.
\usepackage{caption}
\usepackage{subcaption}
\usepackage[table]{xcolor}
\usepackage[section]{placeins}

\begin{document}

\title{Uncertainties from metrology in the integrated luminosity measurement with the updated design of a detector at CEPC}

%%%% To generate auto affiliation numbers please use \author{}\affil{} command

\author[1]{Ivan Smiljanić}
\affil{Vinca Institute of Nuclear Sciences - National Institute of the Republic of Serbia, University of Belgrade, M. Petrovica Alasa 12-14, Belgrade, Serbia \email{i.smiljanic@vin.bg.ac.rs}}

\author[1]{Ivanka Božović}
%\affil{Vinca Institute of Nuclear Sciences - National Institute of the Republic of Serbia, University of Belgrade, M. Petrovica Alasa 12-14, Belgrade, Serbia}

\author[1]{Goran Kačarević}
%\author[3]{Insert fourth author name here} %%% Use optional bracket [3] to change the respective address
%\affil{Vinca Institute of Nuclear Sciences - National Institute of the Republic of Serbia, University of Belgrade, M. Petrovica Alasa 12-14, Belgrade, Serbia}

%\author{Insert last author name here\thanks{These authors contributed equally to this work}}
%\affil{Insert last author address here}

%%% To include the collaborator name... Please use the command "\collaborator"
%%% For example: \collaborator{ATLAS Collaboration}

\begin{abstract}%
In order to measure integrated luminosity with a required precision of $10^{-4}$ at the $Z^0$ pole, proposed CEPC $e^{+}e^{-}$ collider requires a luminometer, a specially designed calorimeter placed in the very forward region to identify Bhabha scattering at low polar angles. Usually, such a device is placed at the outgoing beams, to keep the spatial symmetries of the head-on collisions at accelerators with a non-zero crossing angle. At CEPC it is currently proposed to place the luminometer on the z-axis. We review a feasibility of a measurement of the integrated luminosity at the $Z^{0}$ pole with the required precision, concerning the luminometer centered around the z-axis and the post-CDR beam properties.
\end{abstract}

%\subjectindex{xxxx, xxx}

\maketitle

\section{Introduction}
\label{sec:intro}

In order to achieve precision goals of the electroweak physics program at CEPC at the $Z^{0}$ pole, integrated luminosity $\mathcal{L}$ should be known with the total relative uncertainty of order of $10^{-4}$ \cite{CEPC_CDR}. In the CEPC Conceptual Design Report \cite{CEPC_CDR}, beam properties and detector concept are discussed, with particular emphasis on the machine-detector interface (MDI) and integrated luminosity measurement. The current design of MDI at CEPC has a distinctive feature with respect to the proposal discussed in \cite{CEPC_CDR} - the luminometer is proposed to be placed at the z-axis. This is mainly due to the fact that both detector and machine instrumentation occupy low polar angles, with accelerator components being placed in the detector in a conical space inside the opening angle of $\sim$  118 mrad ($\mid$cos$\theta$$\mid$ $\geq$ 0.993). Although the impact of metrology on the integrated luminosity measurement at CEPC with the luminometer conventionally placed at the outgoing beams (s-axis) has been discussed in \cite{JINST}, it requires revisiting for the newly proposed geometry, since the integrated luminosity precision critically depends on the luminometer's position, as well as on the event selection.

In this paper, we review the potential impact of positioning the luminometer around the z-axis on systematic uncertainties associated with measuring integrated luminosity. These uncertainties stem from mechanical factors like positioning and alignment of the luminometer, as well as from the fact that beam properties are known within some margins. These effects are collectively referred to as metrology. Additionally, we examined the performance of different Bhabha counting methods, including the LEP-style asymmetric counting \cite{LEP1}, typically utilized to mitigate systematic biases arising from asymmetries between the left and right detector halves.

The paper is organized as follows: in Section \ref{sec:sec2} we discuss the current layout of the very forward region of CEPC. In Section \ref{sec:sec3} we review the impact of detector manufacturing, positioning and alignment (Section \ref{sec:sec3.1}), as well as the impact of uncertainties of the beam properties (asymmetry of beam energies, beam energy spread - BES, IP position and beam synchronization) in Section \ref{sec:sec3.2}. The concluding discussion is given in Section \ref{sec:sec4}.

\section{Forward region at CEPC}
\label{sec:sec2}

\begin{figure}[h]
\centering % \begin{center}/\end{center} takes some additional vertical space
\includegraphics[width=.50\textwidth]{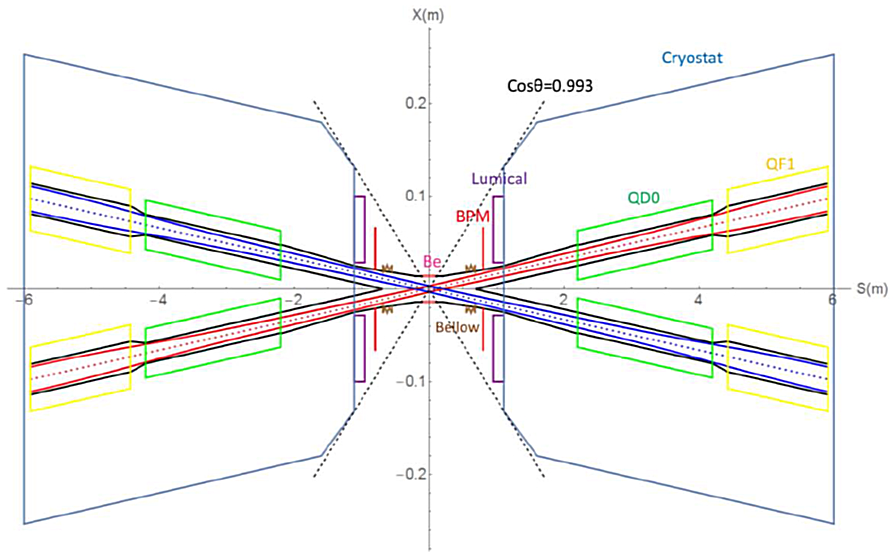}
%\qquad
%\includegraphics[width=.45\textwidth]{Fig3Rf.png}
% "\includegraphics" from the "graphicx" permits to crop (trim+clip)
% and rotate (angle) and image (and much more)
\caption{\label{fig:MDI} Schematic view of the MDI region at CEPC.}
\end{figure}

\begin{figure}[h]
\centering % \begin{center}/\end{center} takes some additional vertical space
\includegraphics[width=.99\textwidth]{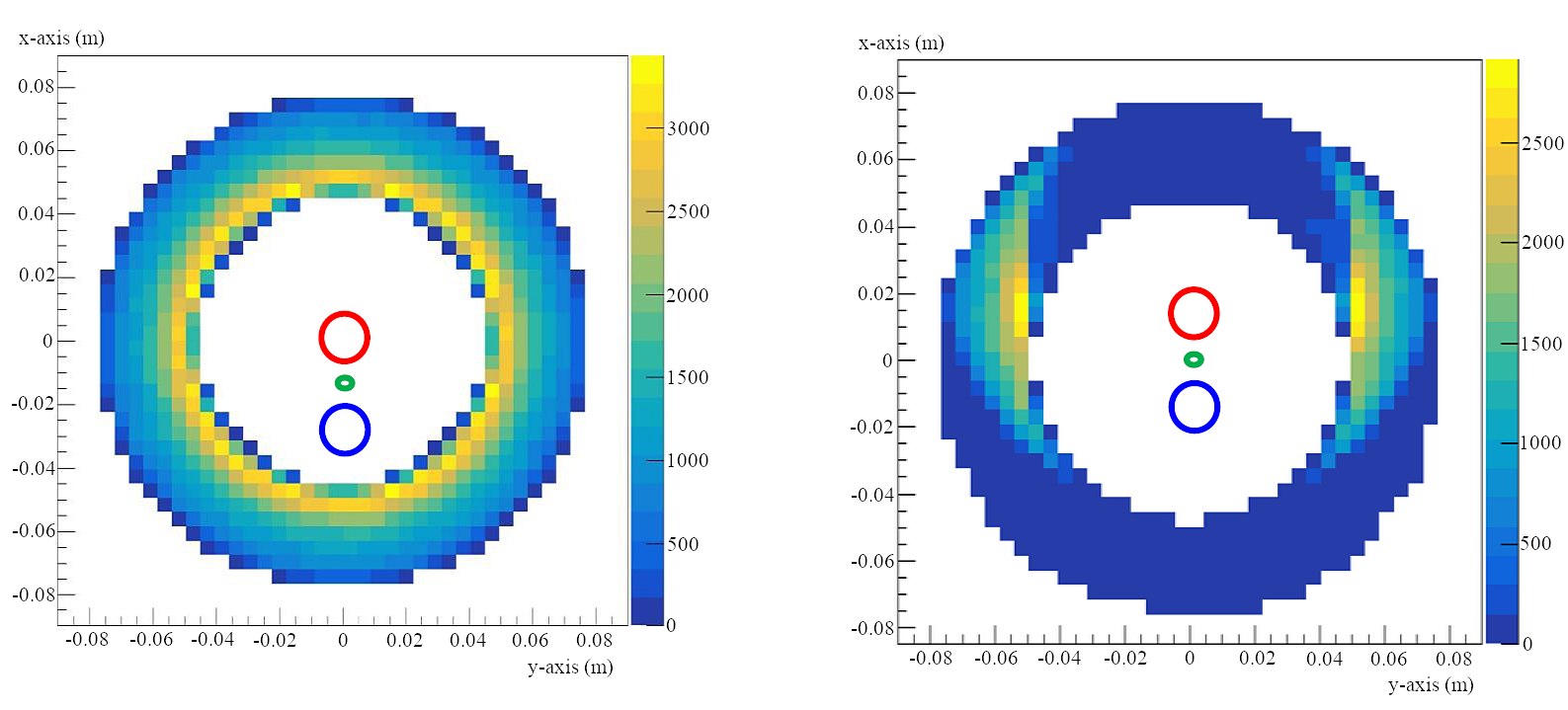}
%\qquad
%\includegraphics[width=.45\textwidth]{Fig3Rf.png}
% "\includegraphics" from the "graphicx" permits to crop (trim+clip)
% and rotate (angle) and image (and much more)
\caption{\label{fig:2} Hitmap of the first plane of the luminometer placed around the outgoing beam (left) and z-axis (right). Red and blue circles represent the outgoing and incoming beam pipes respectively, and the green circle represents the z-axis.}
\end{figure}

The MDI region of CEPC is proposed to longitudinally cover the area of 12 m with the centrally positioned interaction point (IP). The components of the accelerator (without shielding) will be placed in an 118 mrad cone inside the detector, as illustrated in Fig. \ref{fig:MDI} \cite{Hou}. The crossing angle in the horizontal plane between the colliding beams at the IP is 33 mrad. The final focus length of colliding beams is 220 cm \cite{CEPC_CDR}. It is proposed that the luminometer at CEPC should be placed at 95 cm distance from the interaction point, covering the polar angle region from 30 mrad to 105 mrad, which corresponds to the luminometer aperture of 2.85 cm for the inner and 10.0 cm for the outer radius. The fiducial volume of luminometer should be between 53 mrad and 79 mrad. To prevent the leakage of the electromagnetic showers outside of the luminometer’s edges, an iron shielding of 5 mm thickness can be placed around the luminometer. However, this has been done under assumption that the luminometer will be placed around the outgoing beams and it is yet to be proven for the luminometer placed around the z-axis, since the symmetry of signal hits with respect to the detector axis will be lost, as illustrated in Fig. \ref{fig:2} where the hitmap in the first plane of the luminometer is presented. 

The choice of technology of the luminometer at CEPC has not been finally determined yet. However, since this study is based on simulation of counting of low-angle Bhabha scattering without any other assumption but the angular acceptance of the luminometer, it is of no relevance for this study. If one would perform a more realistic simulation including detector response to signal and background, at least an energy cut on electron and positron tracks should be required to separate Bhabha events from the background. In such a case, effects like bias and resolution of energy measurement as well as the uncertainties from calibration of the luminometer should be included as additional sources of systematic uncertainty. The same holds if one includes other selection criteria based on polar and azimuthal angle of Bhabha candidates.

%This demo file is intended to serve as a ``starter file''
%for ptephy journal papers produced under \LaTeX\ using
%\verb+ptephy_v1.cls+ v0.1

\section{Integrated luminosity measurement and systematic uncertainties}
\label{sec:sec3}

Integrated luminosity measurement is conventionally based on the low angle Bhabha scattering (LABS). It is defined as:  

\begin{equation}\label{1.1}
\mathcal L=\frac{N_{Bh}}{\sigma_{Bh}},
\end{equation}

where $N_{Bh}$ is Bhabha count in the certain phase space of parameters and $\sigma_{Bh}$ is the LABS cross-section. Both Bhabha count and theoretical cross-section at the $Z^{0}$ pole should be known at the level of $10^{-4}$, in order to know the integrated luminosity with the same precision. The Bhabha cross-section scales with the center-of-mass energy as \cite{FCAL_JINST}:

\begin{equation}\label{1.2}
\frac{d\sigma}{d\theta} =\frac{2\pi\alpha^2_{em}}{s}\cdot\frac{sin\theta}{sin^4(\frac{\theta}{2})}\approx\frac{32\pi\alpha^2_{em}}{s}\cdot\frac{1}{\theta^3},
\end{equation}

where $\alpha_{em}$ is the QED constant and $\theta$ the polar angle of emitted Bhabha particles. Uncertainty of $\mathrm{10^{-4}}$ of the cross-section implies that the available center-of-mass energy should be known with the absolute uncertainty of $\Delta(\sqrt{s})\leq$ 5 MeV. There is an ongoing study \cite{nesto} at future Higgs factories to check the  feasibility of $\sqrt{s}$ determination using di-muon production reconstructed in the central tracker. However, the required precision seems to be beyond the experimental reach, but, as discussed in \cite{Stahl}, many s-channel processes far from the resonance will have the same dependence of the cross-section on the center-of-mass energy, leading to effective cancellation of the uncertainties in the cross-section measurements. It is up to each individual cross-section measurement analysis to determine if uncertainty of the integrated luminosity measurement caused by the uncertainty of $\sqrt{s}$ is an issue.

\begin{figure}[t]
\centering % \begin{center}/\end{center} takes some additional vertical space
\includegraphics[width=.99\textwidth]{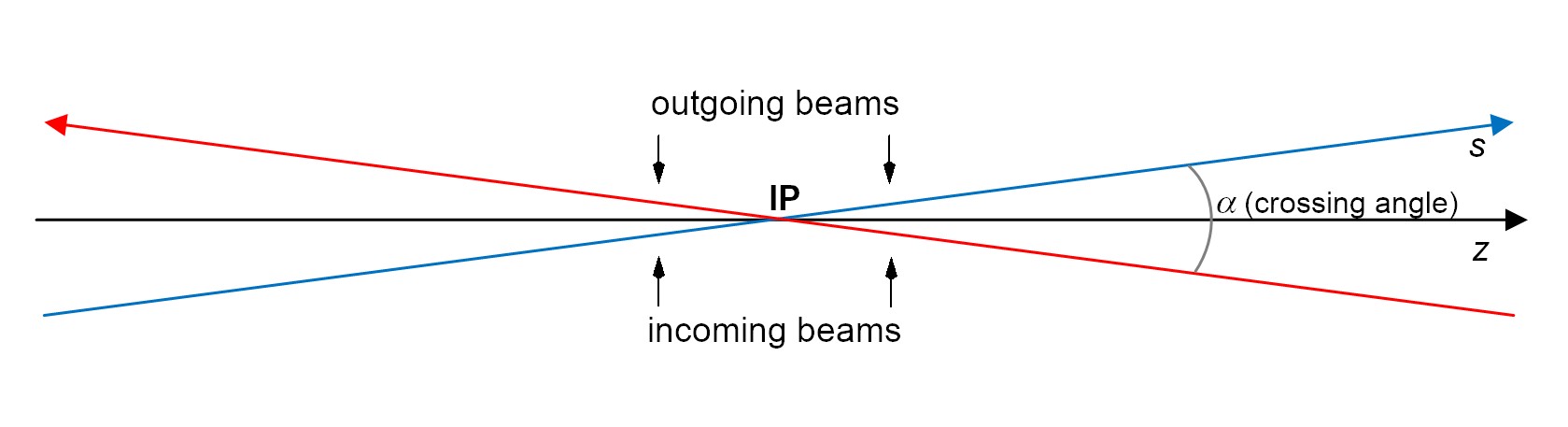}
%\qquad
%\includegraphics[width=.45\textwidth]{Fig3Rf.png}
% "\includegraphics" from the "graphicx" permits to crop (trim+clip)
% and rotate (angle) and image (and much more)
\caption{\label{fig:1} Illustration of s-axis and z-axis.}
\end{figure}

Placement of the luminometer on the z-axis with the CEPC crossing angle of 33 mrad leads to the reduction in Bhabha count for $\sim 70$\% in comparison to the count with the luminometer placed on the s-axis. Fig. \ref{fig:1} provides an illustrative depiction of the s- and z-axes configurations in collisions with the crossing angle. Since the overall Bhabha count in 16 $\mathrm{ab}^{-1}$ of integrated luminosity, which is expected at CEPC at the $Z^{0}$ pole, will be of order of $10^{12}$ for the detector placed on the s-axis, its displacement to the z-axis with the same luminometer apertures will contribute to the relative statistical uncertainty of the count as $\mathrm{1.2 \cdot 10^{-6}}$.

Further, feasibility and requirements for $\mathrm{10^{-4}}$ precision of counting at the $Z^0$ pole will be discussed, considering that the uncertainty of each individual systematic effect contributes to the integrated luminosity at a level of $\mathrm{10^{-4}}$. In \cite{JINST}, at one side of the luminometer the Bhabhas are counted in the full fiducial volume of the luminometer, while at the other side the inner radial acceptance of the fiducial volume is shrunken  by 1 mm. In this study, the outer radial acceptance is also shrunken by 1 mm. This is performed randomly to the left (L) and right (R) sides of the luminometer, on event by event basis, as it has been done at ALEPH \cite{LEP1}. Described way of counting is called asymmetrical (AS) throughout the paper. For symmetric (S) event selection the Bhabhas are counted in  the full fiducial volumes of the both sides of luminometer. As will be discussed further, different detector positioning (s-axis and z-axis) interfere with a way of counting in a nontrivial way. 

\subsection{Uncertainties from mechanics and positioning}
\label{sec:sec3.1}

Systematic uncertainties arising from the detector mechanics and positioning, as well as the uncertainties related to the properties of beams, have been quantified through a simulation study.  We have generated, including the effects of initial (ISR) and final state radiation (FSR),  $2\cdot10^{7}$ Bhabha events scattered at low angles at the $Z^{0}$ pole, using BHLUMI V4.04 Bhabha  event  generator \cite{BHLUMI}. The polar angle range in which the final state particles are generated is from 30 mrad to 100 mrad, to allow events modified by FSR to be detected in the luminometer. In this angular range, the effective Bhabha cross-section is $\sim$ 50 nb.

Considered detector-related uncertainties originating from manufacturing, positioning and alignment of the luminometer are:
\begin{enumerate}
	\item Maximal uncertainty of the luminometer inner aperture, $\Delta r_{in}$,
	\item RMS of the Gaussian dissipation of the radial shower position measured in the luminometer, with respect to the exact position of the Bhabha hit, $\sigma_{r}$ (measured i. e. by placing a tracker plane in front of the luminometer),
	\item RMS of the Gaussian distribution of luminometer fluctuations with respect to the IP, caused by vibrations and thermal stress, in radial, $\sigma_{x_{IP}}$, and axial direction, $\sigma_{z_{IP}}$,
    \item Maximal absolute uncertainty of the distance between left and right sides of the luminometer along the z-axis, $\Delta l$, where both halves are shifted equidistantly for $\pm\frac{\Delta l}{2}$ with respect to the interaction point.
\end{enumerate}

\begin{figure}[htbp]
\centering % \begin{center}/\end{center} takes some additional vertical space
\begin{subfigure}[b]{0.47\textwidth}
\centering
\includegraphics[width=.99\textwidth]{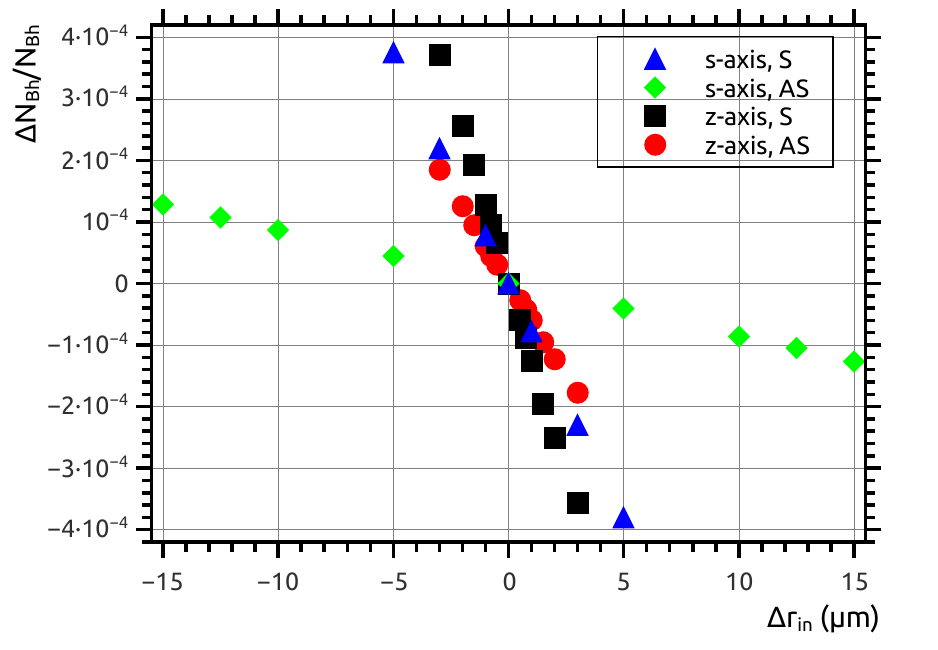}
\caption{}
\end{subfigure}
     \hfill
     \begin{subfigure}[b]{0.47\textwidth}
\centering
\includegraphics[width=.99\textwidth]{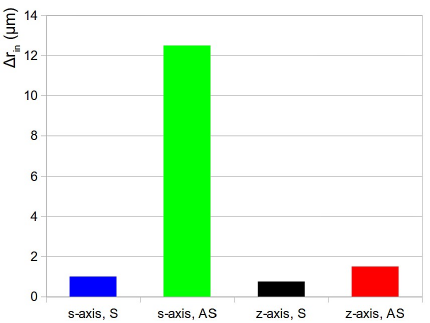}
\caption{}
\end{subfigure}
% "\includegraphics" from the "graphicx" permits to crop (trim+clip)
% and rotate (angle) and image (and much more)
\caption{ (a) Count uncertainty as a function of $\Delta r_{in}$ for two different ways of counting: in the full fiducial volume (S) and the LEP-style (AS), with the detector placed at the s- and z-axis; (b) Variations of the inner aperture of the counting volume, $\Delta r_{in}$, for two different ways of counting (S and AS), for detector placed on the s- and z-axis. Precision limits are taken from the distributions given in (a) for $\Delta N/N$ uncertainty of $10^{-4}$.}
\label{fig:3}
\end{figure}

\begin{figure}[htbp]
\centering % \begin{center}/\end{center} takes some additional vertical space
\begin{subfigure}[b]{0.47\textwidth}
\centering
\includegraphics[width=.99\textwidth]{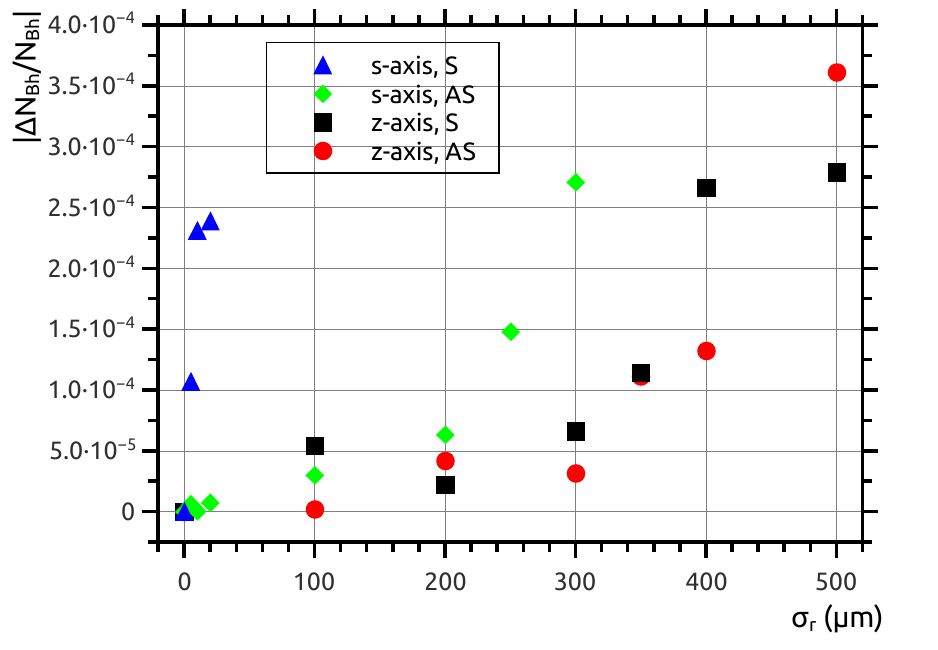}
\caption{}
\end{subfigure}
     \hfill
     \begin{subfigure}[b]{0.47\textwidth}
\centering
\includegraphics[width=.99\textwidth]{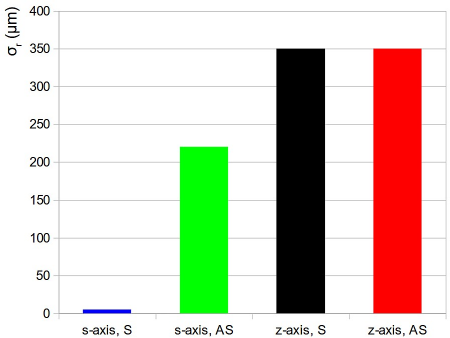}
\caption{}
\end{subfigure}
% "\includegraphics" from the "graphicx" permits to crop (trim+clip)
% and rotate (angle) and image (and much more)
\caption{\label{fig:4} (a) Count uncertainty as a function of $\sigma _ r$ for two different ways of counting (S and AS), with a detector placed at the s- and z-axis. (b) A comparison of results for $\sigma _ r$ for luminometer placed at the s-axis and at the z-axis. Precision limits are taken from the distributions given in (a) for $\Delta N/N$ uncertainty of $10^{-4}$.}
\end{figure}

\begin{figure}[htbp]
\centering % \begin{center}/\end{center} takes some additional vertical space
\begin{subfigure}[b]{0.47\textwidth}
\centering
\includegraphics[width=.99\textwidth]{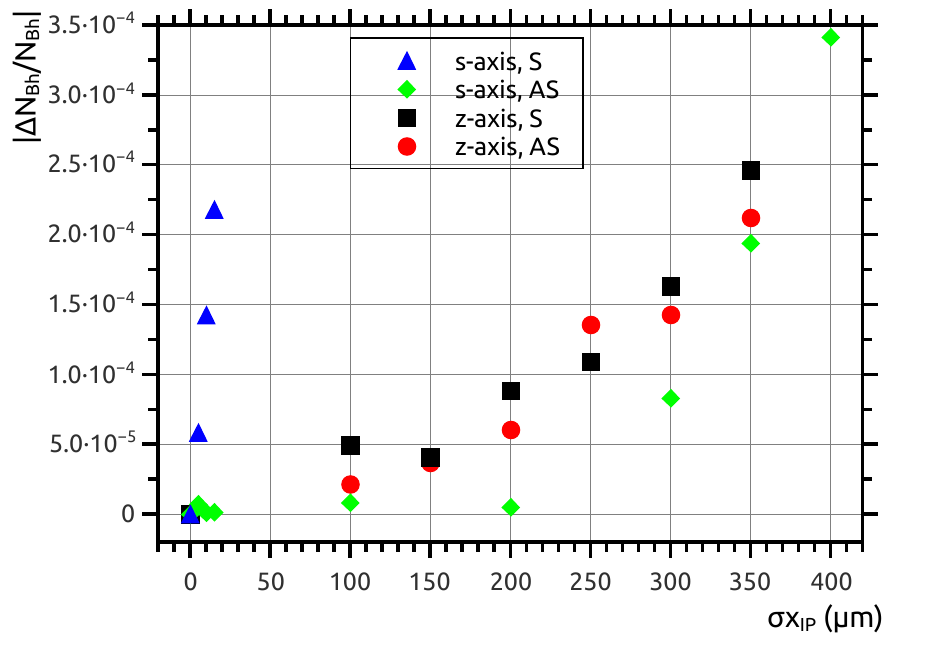}
\caption{}
\end{subfigure}
     \hfill
     \begin{subfigure}[b]{0.47\textwidth}
\centering
\includegraphics[width=.99\textwidth]{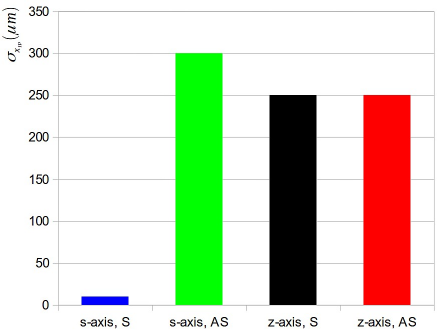}
\caption{}
\end{subfigure}
% "\includegraphics" from the "graphicx" permits to crop (trim+clip)
% and rotate (angle) and image (and much more)
\caption{\label{fig:5} (a) Count uncertainty as a function of $\sigma_{x_{IP}}$ for two different ways of counting (S and AS), with a detector placed at the s- and z-axis; (b) A comparison of results for $\sigma_{x_{IP}}$ for luminometer placed at the s-axis and at the z-axis. Precision limits are taken from the distributions given in (a) for $\Delta N/N$ uncertainty of $10^{-4}$.}
\end{figure}

\begin{figure}[htbp]
\centering % \begin{center}/\end{center} takes some additional vertical space
\begin{subfigure}[b]{0.47\textwidth}
\centering
\includegraphics[width=.99\textwidth]{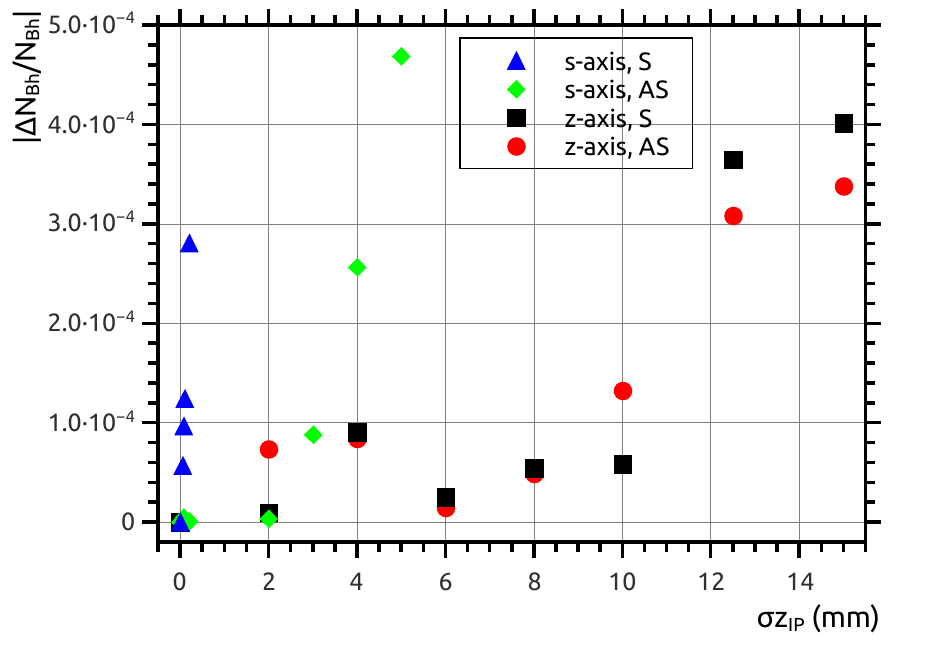}
\caption{}
\end{subfigure}
     \hfill
     \begin{subfigure}[b]{0.47\textwidth}
\centering
\includegraphics[width=.99\textwidth]{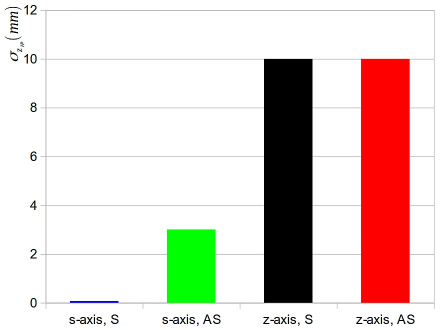}
\caption{}
\end{subfigure}
% "\includegraphics" from the "graphicx" permits to crop (trim+clip)
% and rotate (angle) and image (and much more)
\caption{\label{fig:6} (a) Count uncertainty as a function of $\sigma_{z_{IP}}$ for two different ways of counting (S and AS), with a detector placed at the s- and z-axis; (b) A comparison of results for $\sigma_{z_{IP}}$ for luminometer placed at the s-axis and at the z-axis. Precision limits are taken from the distributions given in (a) for $\Delta N/N$ uncertainty of $10^{-4}$.}
\end{figure}

Figures \ref{fig:3}-\ref{fig:6} illustrate effects 1-4 respectively, for different ways of counting and different positions of the luminometer placed on the s- and z-axis. Values in histograms shown on Figures \ref{fig:3}(b)-\ref{fig:6}(b) are based on the data from Figures \ref{fig:3}(a)-\ref{fig:6}(a), obtained from the local fit in the region of interest. If the distribution on (a) was not symmetrical for positive and negative variations of the parameter, the more restrictive limit was considered.

\begin{figure}[htbp]
\centering % \begin{center}/\end{center} takes some additional vertical space
\begin{subfigure}[b]{0.47\textwidth}
\centering
\includegraphics[width=.99\textwidth]{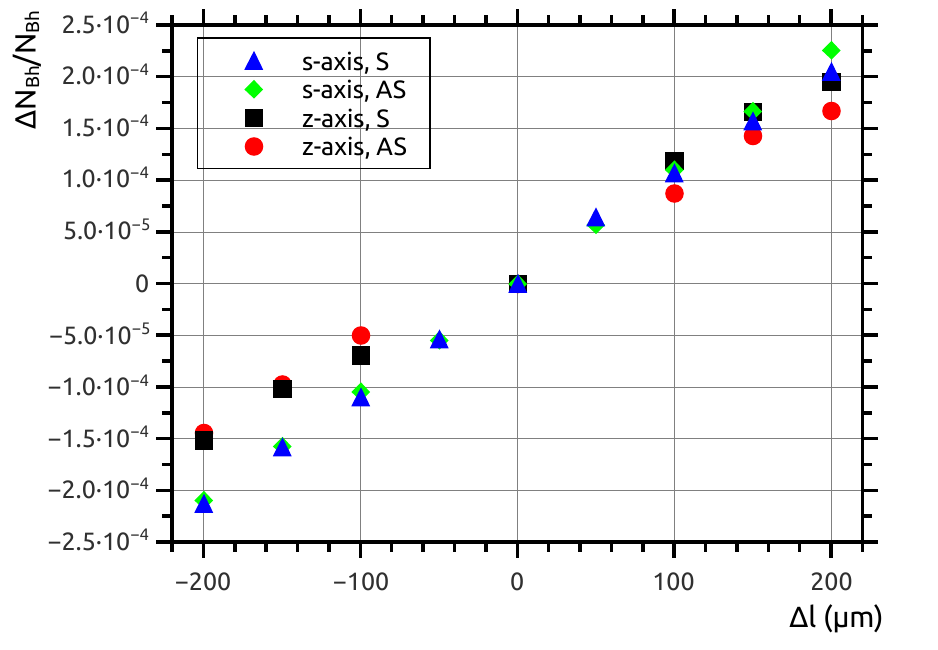}
\caption{}
\end{subfigure}
     \hfill
     \begin{subfigure}[b]{0.47\textwidth}
\centering
\includegraphics[width=.99\textwidth]{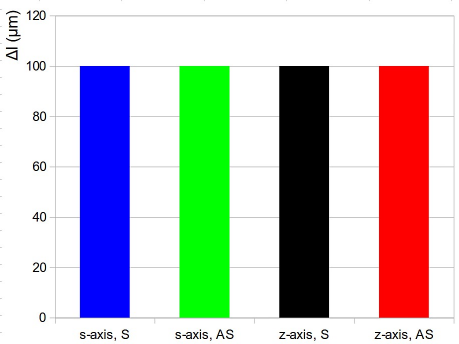}
\caption{}
\end{subfigure}
% "\includegraphics" from the "graphicx" permits to crop (trim+clip)
% and rotate (angle) and image (and much more)
\caption{\label{fig:6a} (a) Count uncertainty as a function of $\Delta l$ for two different ways of counting (S and AS), with a detector placed at the s- and z-axis; (b) A comparison of results for $\Delta l$ for luminometer placed at the s-axis and at the z-axis. Precision limits are taken from the distributions given in (a) for $\Delta N/N$ uncertainty of $10^{-4}$.}
\end{figure}

Figures \ref{fig:3}-\ref{fig:6} show that for the detector placed at the z-axis asymmetric counting performs similarly as counting in the full fiducial volume, whereas  with the detector placed at the s-axis the difference between the two ways of counting is quite significant. In Figure \ref{fig:3}, which illustrates the impact of $\Delta r_{in}$, it can be seen that the requirement on the precision of inner radius of the counting volume is relaxed with the detector placed at the s-axis. This is due to the fact that asymmetric counting allows $\Delta r_{in}$ to modify one side of the luminometer, while the other side of the counting volume is shrunken, compensating for the uncertainty $\Delta r_{in}$. This compensation works only if signal hits are L-R symmetrical, which is the case for the detector placed at the s-axis, and effect itself introduces L-R asymmetry. When the detector is placed at the z-axis, the symmetry of signal hits is lost (Figure \ref{fig:2}) and, therefore, in this case asymmetric counting is no better than the symmetric counting in the full FV. As illustrated for $\sigma_{r}$ (Figure \ref{fig:4}), $\sigma_{x_{IP}}$ (Figure \ref{fig:5}) and $\sigma_{z_{IP}}$ (Figure \ref{fig:6}), asymmetric counting works significantly better than symmetric for the detector placed at the s-axis and doesn't make a difference for the detector placed at the z-axis. If an effect does not introduce L-R asymmetry, like $\Delta l$ illustrated in Figure \ref{fig:6a}, there would be no difference if counting was performed in the full FV on both sides of the luminometer, or in the asymmetric way. One has to keep in mind that the results shown in this paper are based on distributions of mean values, each obtained from the simulated signal sample of $2 \cdot 10^7$ events, with intrinsic relative statistical uncertainty of $\sim 2\cdot 10^{-4}$. Therefore, the extracted precision limits rather reflect the order of magnitude required to control certain mechanical or beam properties than to be the exact values of detector or beam parameter uncertainties.

%\clearpage

%\clearpage

\begin{figure}[htbp]
\centering % \begin{center}/\end{center} takes some additional vertical space
\begin{subfigure}[b]{0.47\textwidth}
\centering
\includegraphics[width=.99\textwidth]{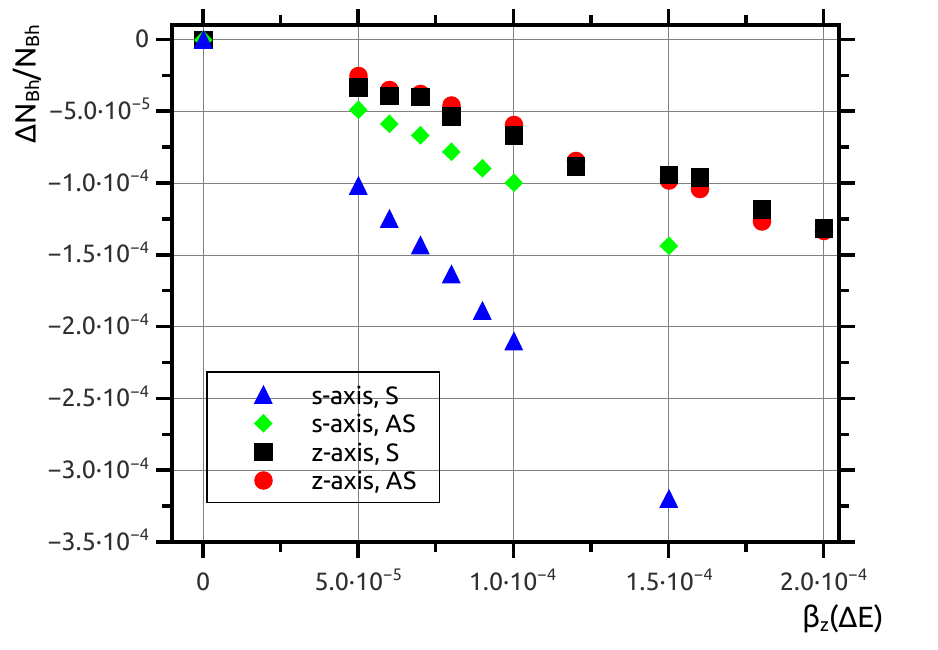}
\caption{}
\end{subfigure}
     \hfill
     \begin{subfigure}[b]{0.47\textwidth}
\centering
\includegraphics[width=.99\textwidth]{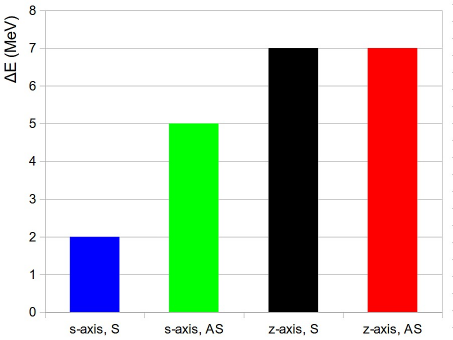}
\caption{}
\end{subfigure}
% "\includegraphics" from the "graphicx" permits to crop (trim+clip)
% and rotate (angle) and image (and much more)
\caption{\label{fig:7} (a) Loss of the Bhabha count in the luminometer due to the longitudinal boost of the center-of-mass frame, $\beta_{z}(\Delta E)$, for two different ways of counting (S and AS), with a detector placed at the s- and z-axis; (b) A comparison of results for $\Delta E$ for luminometer placed at the s-axis and at the z-axis. Precision limits are taken from the distributions given in (a) for $\Delta N/N$ uncertainty of $10^{-4}$.}
\end{figure}

\begin{figure}[htbp]
\centering % \begin{center}/\end{center} takes some additional vertical space
\begin{subfigure}[b]{0.47\textwidth}
\centering
\includegraphics[width=.99\textwidth]{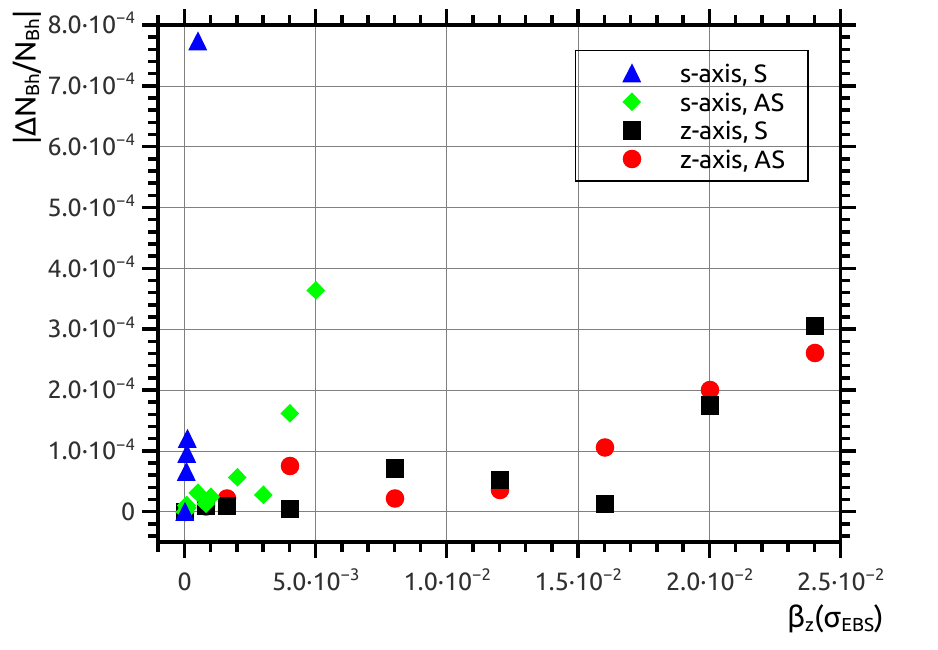}
\caption{}
\end{subfigure}
     \hfill
     \begin{subfigure}[b]{0.47\textwidth}
\centering
\includegraphics[width=.99\textwidth]{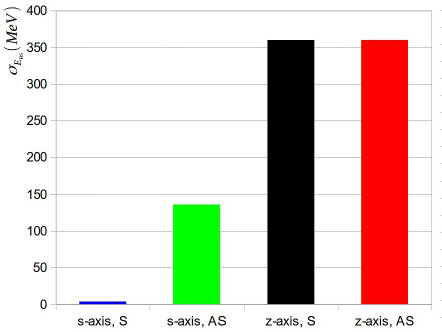}
\caption{}
\end{subfigure}
% "\includegraphics" from the "graphicx" permits to crop (trim+clip)
% and rotate (angle) and image (and much more)
\caption{\label{fig:8} (b) Loss of the Bhabha count in the luminometer due to longitudinal boost of the Bhabha center-of-mass frame caused by BES, $\beta_{z}(\sigma _{E_{BS}})$, for two different ways of counting (S and AS), with a detector placed at the s- and z-axis; (b) A comparison of results for $\sigma _{E_{BS}}$ for luminometer placed at the s-axis and at the z-axis. Precision limits are taken from the distributions given in (a) for $\Delta N/N$ uncertainty of $10^{-4}$.}
\end{figure}

\begin{figure}[htbp]
\centering % \begin{center}/\end{center} takes some additional vertical space
\begin{subfigure}[b]{0.47\textwidth}
\centering
\includegraphics[width=.99\textwidth]{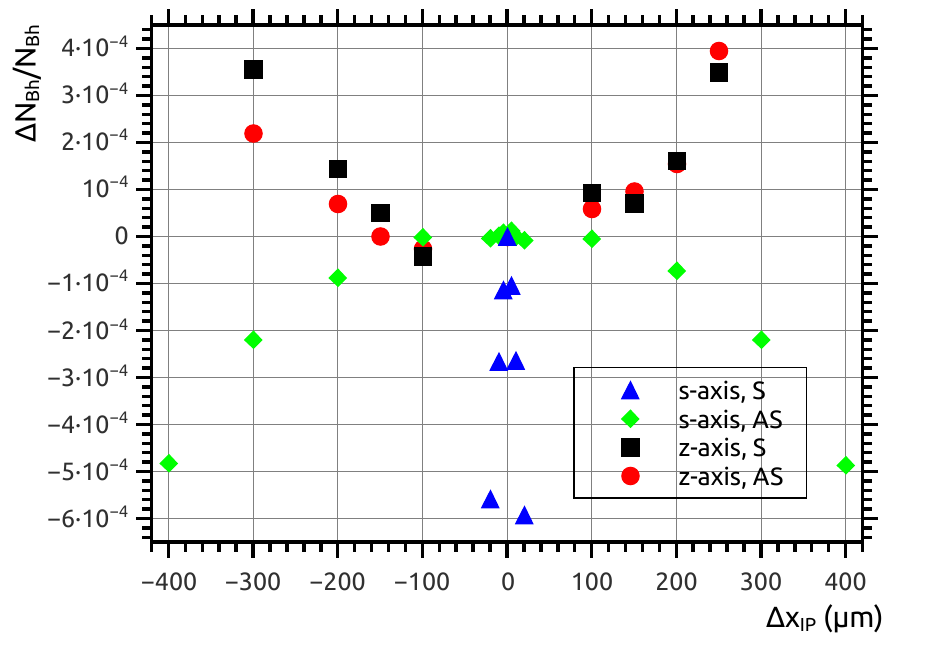}
\caption{}
\end{subfigure}
     \hfill
     \begin{subfigure}[b]{0.47\textwidth}
\centering
\includegraphics[width=.99\textwidth]{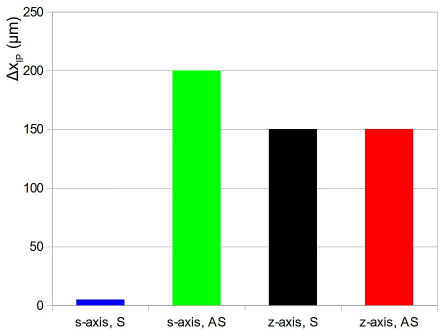}
\caption{}
\end{subfigure}
% "\includegraphics" from the "graphicx" permits to crop (trim+clip)
% and rotate (angle) and image (and much more)
\caption{\label{fig:9} (a) Count uncertainty as a function of $\Delta x_{IP}^{BS}$ for two different ways of counting (S and AS), with a detector placed at the s- and z-axis; (b) A comparison of results for $\Delta x_{IP}^{BS}$ for luminometer placed at the s-axis and at the z-axis. Precision limits are taken from the distributions given in (a) for $\Delta N/N$ uncertainty of $10^{-4}$.}
\end{figure}

\begin{figure}[htbp]
\centering % \begin{center}/\end{center} takes some additional vertical space
\begin{subfigure}[b]{0.47\textwidth}
\centering
\includegraphics[width=.99\textwidth]{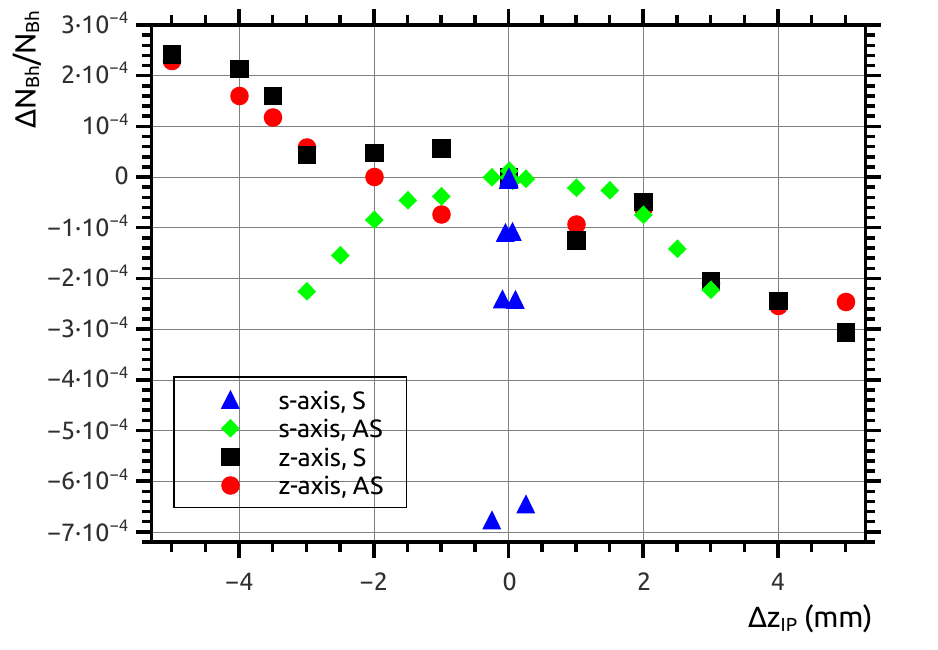}
\caption{}
\end{subfigure}
     \hfill
     \begin{subfigure}[b]{0.47\textwidth}
\centering
\includegraphics[width=.99\textwidth]{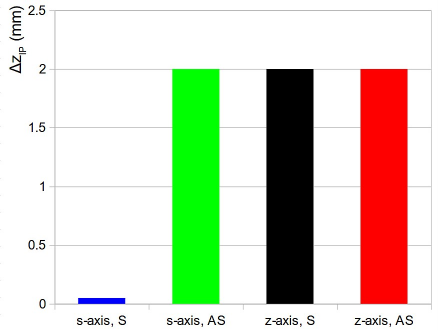}
\caption{}
\end{subfigure}
% "\includegraphics" from the "graphicx" permits to crop (trim+clip)
% and rotate (angle) and image (and much more)
\caption{\label{fig:10} (a) Count uncertainty as a function of $\Delta z_{IP}^{BS}$ for two different ways of counting (S and AS), with a detector placed at the s- and z-axis; (b) A comparison of results for $\Delta z_{IP}^{BS}$ for luminometer placed at the s-axis and at the z-axis. Precision limits are taken from the distributions given in (a) for $\Delta N/N$ uncertainty of $10^{-4}$.}
\end{figure}

\clearpage

\subsection{Beam related uncertainties}
\label{sec:sec3.2}

Several systematic uncertainties may arise from uncertainties of the properties of the beams and their delivery to the interaction point. The list of main CEPC beam parameters is given in Table \ref{tab:1} \cite{Gao}.

\begin{table}[h]
	\centering
	\caption{\label{tab:1} CEPC TDR beam parameters at the $Z^0$ pole.\\}
%\small
	\begin{tabular}{|l|c|}
	\hline 
	\textbf{parameter}  & \textbf{$Z^0$ pole} \\ 
		\hline 
            Half crossing angle at IP (mrad) & 16.5 \\ 
	    \hline 
	    Energy (GeV) & 45.5 \\ 
	    \hline 
	    Bunch population ($\mathrm {10^{11}}$) & 1.4 \\ 
	    \hline 
            Bunch length (natural/total) (mm) & 2.7/10.6 \\
            \hline
	    Beam size at IP $\sigma_x/\sigma_y$ ($\mathrm{\mu m/nm}$) & 6/35 \\ 
	    \hline 
	    Energy spread (natural) (\%) & 0.04 \\ 
	    \hline 
		Luminosity per IP ($\mathrm{10^{34}}$ $\mathrm{cm^{–2} s^{–1}}$) & 115 \\ 
            \hline
\end{tabular} 
\end{table}

%\clearpage

Here we consider:

\begin{enumerate}
        \item maximal permanent bias ($\Delta E$) of a single beam energy with respect to the other beam, resulting in a longitudinal boost of the incoming $e^+ e^-$ system with respect to the laboratory frame, and consequently, boost of the Bhabha center-of-mass system,
        \item maximal RMS of the Gaussian distribution of the beam energy spread ($\sigma _{E_{BS}}$), responsible for longitudinal boost on event-by-event basis, leading to the overall count loss of order of $\mathrm{10^{-4}}$,
	\item maximal radial  ($\Delta x_{IP}^{BS}$) and axial  ($\Delta z_{IP}^{SY}$) IP position displacements with respect to the luminometer arms, caused by the finite beam sizes (former) and beam synchronization (latter),
	\item maximal time shift in beam synchronization ($\Delta \tau_{SY}$), causing the IP longitudinal displacement  $\Delta z_{IP}^{SY}$. 
\end{enumerate}

\clearpage

In the same manner as in the previous chapter, results for effects 1-4 are illustrated in Figures \ref{fig:7}-\ref{fig:10} and the upper limits, together with the upper limits of mechanical parameters from Section \ref{sec:sec3.1}, are summarized in Table \ref{tab:2}. All limits of parameters given in Table \ref{tab:2} are derived from distributions of mean values in Figures \ref{fig:3}(a) - \ref{fig:10}(a), without considering statistical uncertainties from the limited size of the simulated samples. Each effect is considered to individually contribute to the the relative uncertainty of the integrated luminosity as $1 \cdot 10^{-4}$.

\begin{table}[]
	\centering
	\caption{\label{tab:2} Minimal absolute precision of luminometer mechanical parameters and beam parameters, each contributing as $10^{-4}$ to the relative uncertainty of $\mathcal L$ at the $Z^{0}$ pole. The average net center-of-mass energy uncertainty $\Delta E_{CM}$ limit is derived by error propagation from the Bhabha cross-section dependence on the center-of-mass energy.\\}
	\small
	\begin{tabular}{|l|c|c|c|c|}
	\hline 
	\textbf{parameter}  & \textbf{s-axis, symm.} & \textbf{s-axis, asymm.} & \textbf{z-axis, symm.} & \textbf{z-axis, asymm.}\\ 
		\hline 
            $\Delta r_{in}$ ($\mu$m) & 1 & 12 & 0.8 & 1.5 \\ 
	    \hline 
	    $\sigma_{r}$ ($\mu$m) & 5 & 220 & 350 & 350 \\ 
	    \hline 
	    $\sigma_{x_{IP}}$ ($\mu$m) & 10 & 300 & 250 & 250 \\ 
	    \hline 
	    $\sigma_{z_{IP}}$ (mm) & 0.08 & 3 & 10 & 10 \\ 
	    \hline 
        $\Delta l$ ($\mu$m) & 100 & 100 & 100 & 100 \\ 
	    \hline
		$\Delta E_{CM}$ (MeV) & 5 & 5 & 5 & 5\\ 
		\hline 
		$\Delta E$ (MeV) & 2 & 5 & 7 & 7 \\ 
		\hline 
            $\sigma _{E_{BS}}$ (MeV) & 4 & 140 & 360 & 360 \\
            \hline
		$\Delta x_{IP}^{BS}$ ($\mu$m) & 5 & 200 & 150 & 150 \\ 
		\hline 
		$\Delta z_{IP}^{SY}$ (mm) & 0.05 & 2 & 2 & 2 \\ 
		\hline 
		$\Delta \tau$ (ps) & 0.2 & 7 & 7 & 7 \\ 
		\hline 
\end{tabular} 
\end{table}

As can be seen from Table \ref{tab:2}, mechanical precision of the inner aperture of the counting volume seems to be the major challenge, since the Bhabha cross-section is scaling with the polar angle as $\sigma \sim 1/\theta^3$. However, asymmetric counting applicable at the s-axis relaxes this requirement, while for the z-axis the inner radius of the counting volume should be controlled with precision $\leqslant 1$ $\mathrm{\mu m}$. The longitudinal boost $\delta E$ of the Bhabha center-of-mass frame with respect to the laboratory frame, $\beta_{z}$, may be caused by any difference in beam energies ($\beta_{z}=2 \cdot \delta E/\sqrt{s}$), either as a bias of energy of one beam with respect to the other, $\Delta E$,  or as a random asymmetry in beam energies caused by the beam energy spread or by other radiative losses. While the beam energy spread at CEPC is not an issue for integrated luminosity measurement at the $Z^{0}$ pole, bias in beam energies, $\Delta E$, can be tolerated up to $\mathrm{7 \: MeV}$ ($1.5 \cdot 10^{-4}$ of the nominal beam energy). From Table \ref{tab:2} it can be seen that  displacement of the luminometer from the s-axis to the z-axis will not impact the luminosity precision significantly, while some dependencies will be modified due to loss of spacial symmetries of Bhabha events.

\section{Conclusion}
\label{sec:sec4}
Although the method of integrated luminosity has been studied in a great detail at LEP (\cite{LEP1}, \cite{LEP2}), for each newly proposed $e^{+}e^{-}$ collider it is necessary to quantify the achievable luminosity precision, in particular at the $Z^{0}$ resonance. Here it is done for CEPC, taking into consideration the mechanical and beam-related requirements at the $Z^{0}$ pole, with the luminometer placed on the z-axis. It is the first attempt to quantify systematic effects rising from metrology with the luminometer displaced from the the outgoing beams, where it is conventionally positioned at colliders with a crossing angle.

Control of the luminometer inner radius at the micrometer level at the $Z^{0}$ pole seeems to be the most demanding requirement for the detector manufacturing. Any additional change in the luminometer design towards smaller polar angle coverage will require control of the luminometer inner radius even below micrometer precision, according to the dependence of the Bhabha cross-section on the polar angles of the scattered particles. However, it is important to note that these results are obtained under assumption that the change of the luminometer's inner radius corresponds to exactly the same change of the inner radius of the luminometer's fiducial volume, which needs to be approved. Also, control of the asymmetrical bias in beam energies might be challenging ($1.5 \cdot 10^{-4}$), while the beam energy spread can be relaxed significantly beyond the current RMS of 18 MeV at the $Z^{0}$ pole. Existing techniques like the Frequency Scanning Interferometry (FSI) allow positioning control of the luminometer at the micrometer level \cite{Daniluk}, thus the mechanical effects from metrology should not be issue for the integrated luminosity measurement.

From the point of view of experimental input for the Bhabha cross-section calculations, center-of-mass energy should be known at the $\sim{10^{-4}}$ level, which requires further studies in terms of feasibility of such a precision. This might be an issue for the cross-section measurements on the slope of a production threshold.

Although the simulated Bhabha samples are relatively small to be exact on boundaries of the considered systematics, distribution of central values presented in this study confirms that no effect seems to be more critical for precision measurement of the integrated luminosity if the luminometer is placed around the z-axis than if it is placed around the outgoing beams, except for the inner aperture of the counting volume that can be relaxed with the asymmetric counting at the s-axis. The LEP-style asymmetric counting, however, will not be effective for the luminometer placed on the z-axis to cope with the effects arising from the left-right asymmetry.

Complex metrology, together with the beam-related interactions affecting the Bhabha final state at small polar angles \cite{Voutsinas} may call for consideration of an alternative or a complementary central process for the integrated luminosity measurement, like di-photon or di-muon production (under the assumption that this process at the $Z^{0}$ pole energy doesn't receive relevant contributions from the processes beyond the Standard Model), with hopefully less complex systematic effects subjected to further studies \cite{ECFA}. In addition, studies presented here should be complemented with a full simulation study of systematic effects arising from the detector design, technology and performance in the presence of realistic backgrounds.

\section*{Acknowledgment}

This research was funded by the Ministry of Education, Science and Technological Development of the Republic of Serbia and by the Science Fund of the Republic of Serbia through the Grant No. 7699827, IDEJE HIGHTONE-P.


%\bibliographystyle{ptephy}
%\bibliography{sample}
%
% once the .bbl file has been generated then place the text in your article.

\vspace{0.2cm}
\noindent
%For references,  note how to include DOI information from examples below. 

%This is added by T. Yoneya (editor-in-chief) on 2020/07/09.

\let\doi\relax

%without this code before the command "\begin{thebibliography}{}" , an error will be %flagged. When the bibliography is provided as separate .bib file, then this code %should be placed above the commands "\bibliographystyle{}" and "\bibliography{}" %inside the main TeX file. 

\end{document}